# A diagnosis of the primary difference between EuroForMix and STRmix™

John Buckleton[1,2], Mateusz Susik[3,4], James M. Curran[2], Kevin Cheng[1], Duncan Taylor[5,6], Jo-Anne Bright[1], Hannah Kelly[1], Richard Wivell[1].

1. Institute of Environmental Science and Research Limited, Private Bag 92021, Auckland, 1142 New Zealand

2. University of Auckland, Department of Statistics, Private Bag 92019, Auckland 1142, New Zealand

3. Biotype GmbH, Dresden, 01109, Germany

4. Technische Universit¨ at Dresden, Faculty of Computer Science, Dresden,01187, Germany

5. Forensic Science SA, GPO Box 2790, Adelaide, SA 5001, Australia

6. School of Biological Sciences, Flinders University, GPO Box 2100 Adelaide SA, Australia 5001

**Acknowledgements**

This work was supported in part by grant NIJ 2020-DQ-BX-0022 from the US National Institute of Justice. Points of view in this document are those of the authors and do not necessarily represent the official position or policies of their organizations. We thank Adam McCarthy for comments that greatly improved this paper.

**Conflict of interest**

Buckleton, Taylor and Bright are the developers of STRmix™. However they have no financial interest in the software.

Susik is the developer of HMS but also has no financial interest in the product.

**Abstract**

There is interest in comparing the output, principally the likelihood ratio, from the two probabilistic genotyping software EuroForMix (EFM) and STRmix™. Many of these comparison studies are descriptive and make little or no effort to diagnose the cause of difference. There are fundamental differences between EFM and STRmix™ that are causative of the largest set of likelihood ratio differences. This set of differences is for false donors where there are many instances of *LR*s just above or below 1 for EFM that give much lower *LR*s in STRmix™. This is caused by the separate estimation of parameters such as allele height variance and mixture proportion under $H_p$ and $H_a$ for EFM. It results in a departure from calibration for EFM in the region of *LR*s just above and below 1.

**Highlights**

- Meester and Slooten inform the main difference between EuroForMix and STRmix™
- Separate optimization of numerator and denominator causes a non-calibration for EuroForMix
- Empirical evidence supports this

# 1.0 INTRODUCTION

Continuous DNA interpretation systems utilise peak height, molecular weight, and allelic designation to assign a probability to some observed evidence profile(s) given a potential contributing genotype set. Such systems involve computation of complex integrals which are generally not tractable through hand-calculation, and therefore these calculations are all embedded in software, termed probabilistic genotyping (PG) software. There is considerable, and valid, interest in the reliability of these software products.

Reliability is not easy to define, and we make no meaningful effort to do so here. The key to the discussion we seek to open here is the concept of comparing two independent assignments of an *LR* from two different PG models. If these are similar, then there is at least some possibility that both are accurate. If they differ substantially, then this sheds doubt on at least one and maybe both systems. However, noting that two software produce different answers does not inform which, if any, of the two is reliable. Equally noting that they have different models is no help to the forensic scientist reporting in court. The experts should know which, if any, of the two software is reliable. This can only be undertaken by calibration [1] and is not meaningfully informed by comparison of the software.

We will use the terms $H_p$ and $H_a$ (sometimes $H_d$) for true and false donor tests, respectively.

In June 2021 a team of authors at the National Institute of Standards and Technology (NIST) published a document for public comment that is referred to as the *draft NIST foundation review* [2]. This document states:

> *"Different experts using different assumptions, different statistical models, and different inference procedures may arrive at different LR values. Information regarding the extent to which their LR values agree or disagree is typically not available. There appears to be a general misconception that LR assessments made by different experts will be close enough to one another to not materially affect the outcome of a case. Although they may be close enough in many instances, this is not known for any particular case and it is not advisable to take this for granted."*

Later in the same NIST foundation review:

> *"Since repeatability and reproducibility are components of reliability, it is fair to ask to what extent the LR values offered by different experts using different databases and different models differ from one another. If the accuracy and reliability of a specific LR assignment is important to a case, then understanding what level of reproducibility there is between laboratories or between forensic scientists will help assess reliability. Whereas each laboratory or expert may feel justified in considering their assessments to be reliable, the recipients of such assessments in a given case need guidance on what to do in situations where variation among different LR assessments could impact the outcome of a trial. In particular, because there are no standards to compare to and no traceability considerations as there are for measurements, judgments of reliability by decision makers or triers of fact will be helped by comparing LR assessments from multiple systems and made by multiple experts."*

This discussion has also occurred in court. In the ruling denying admissibility of the PG software STRmix™ for Wisconsin v Troy Williams the Hon. Carolina Stark [3] criticised the lack of comparison of results given different PG software as a test of reliability of the software. It has also been raised in scientific publications which compare the output of multiple PG software in interpretations [4] (and see [5] for comment).

Recently [6] the German project group "Biostatistical DNA Calculations" and the German Stain Commission gave recommendations for the reporting of the output of Probabilistic Genotyping Systems. Their recommendations are affected by an observed four orders of magnitude difference in the *LR*s produced for the same samples by four different software. They have advised that any *LR* less than $10^6$ be described as inconclusive, partly because of these differences.

Given these statements, it is therefore vital to diagnose the cause of these differences and that is the intent of this paper.

We have been aware of a major difference in performance between two commonly used PG software, EuroForMix (EFM) and STRmix™ for some time. This difference occurs for non-donors who produce *LR*s relatively close to, but either side of, 1 within EFM but much lower *LR*s within STRmix™. This is the most observable feature of *x-y* scatter plots of *LR*s for non-donors for the two software. It occurs across a range of versions of either software. This difference can be seen in Figure 1 (the area enclosed by the dotted rectangle), and further evidence is presented later in this paper.

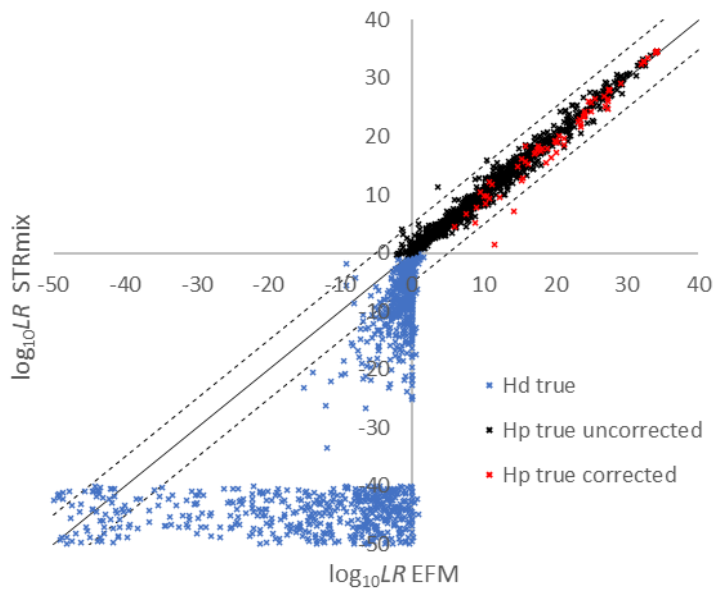

Figure 1: The data from Riman et al. [7] collated from their supplementary material. The corrected data refers to reruns of EFM using V3.2.0 that can handle the type of input file used and two corrections to the STRmix™ runs due to a primer binding site mutation and very poor PCR. The reruns replace the original data. Reruns of the EFM value are not available for all data. The $x = y$ and $x = y \pm 5$ lines are given. Exclusions (*LR*=0) are plotted randomly between $\log_{10}LR$=-40 to -50. The dotted triangle encloses the area with the differences that are the primary subject of this publication.

More recently Meester and Slooten [8] published a book on probability and forensic evidence. This book is insightful but heavily mathematical. It does not seem to, as yet, have had the impact on the forensic community that is deserved. It contains a section entitled "Maximum likelihood versus integration." This section is the clue to explaining the difference that we have observed between the two software. It suggests that this difference is fundamental to the process of the two software and unlikely to be changed by any small changes to the models. The difference is that EFM in most uses applies Maximum Likelihood Estimation (MLE) and STRmix™ always applies integration (using MCMC).

Benschop et al. [9], who use EFM, report that the Netherlands Forensic Institute do not report *LR*s less than $10^4$ "*as calculations resulting in lower LRs can be more sensitive to the model and parameters used.*" It is essential that any exclusionary evidence is reported. We hypothesise that it may be motivated by EFMs tendency to assign too high *LR*s for non-donors.

In this paper we reprise the evidence from previous comparisons often reprocessing the output into simpler *x-y* scatter plots. Where possible we amend faults in the comparison studies or highlight the existence of such faults. These faults tend to create outliers but do not impact the large differences noted above. We add data from a Hamiltonian Monte Carlo method (HMC) [10, 11] which is a highly credible method with a near clone of the biological models used in STRmix™, but applied through a different stochastic sampling scheme. Expectedly *LR*s from this method tend to plot near those assigned by STRmix™ for the same inputs. This is further evidence that the difference is based on fundamentals of the software and not the details of implementation. This paper combines existing and new data to research and confirm the insight that Meester and Slooten [8] provided; that the use of MLE and conditioning separately in the numerator and denominator will assign too high *LR*s for non-donors.

### 1.1 *Analysis*

Riman et al. [7] state: "*… depending on the software being used, the analysis of the same DNA profile could yield different numeric LR values and, if used, different verbal characterization.*" The Riman et al. paper compares the numerical outputs of STRmix™ (V2.6) and EFM (V2.1.0 for most data and V3.2.0 for the rerun of a limited subset) for a wide range of mixtures.

They demonstrate that the *LR* results obtained from multiple samples using the two software are similar but do report some notable differences for some samples. In Figure 1**Error! Reference source not found.** we reproduce their assigned *LR*s for STRmix™ and EFM for both $H_p$ true tests (true donors) and $H_a$ true tests (non-donors).

The conclusions of this potentially valuable study suffer from the demonstrated inappropriate application of the two software. The Riman et al. study has a number of flaws noted by both Riman et al. themselves (see for example their Section 2.5 describing the retention of stutter labels in EFM inputs for stutter not explicitly modelled within that software) and Buckleton et al. [12]. In brief summary, they used both software EFM and STRmix™ unvalidated, on input created by a different software CleanIt [13] that they had also not validated by itself or in conjunction with EFM and STRmix™. In addition, they knowingly retained artifacts within the input files that they knew EFM did not model. The CleanIt software removes some artifacts such as some large back stutter, that is modelled within both software and hence needs to be retained [13].

More recently, Susik et al. [11] describes and tests a Hamiltonian Monte Carlo method (HMC) for the continuous interpretation of DNA profiles. This HMC method is a close analogue of the models employed within STRmix™. They also provide comparative results between HMC, STRmix™, and EFM using the NIST MIX05 and MIX13 inter-laboratory study. We reproduce a part of their output for the NIST MIX13 study in **Error! Reference source not found.**.

Table 1: Log$_{10}$*LR*s for HMC, EFM (V1.10.1 and V1.11.4) and STRmix™ (V2.5.11) for the NIST MIX13 interlaboratory study reproduced from Susik et al. [11]. In the propositions 'V' in this list is always a conditioning profile. NoC is the number of contributors. *Note that for EFM and STRmix™ from the original paper [14] the results for reruns varied. The numbers here are quoted from Susik et al. and are the log of the average. They include a correction provided by Susik (Log$_{10}$*LR*s of 6.17 instead of 6.45 for the proposition "S05A+U+U"). Negative infinity represents *LR*=0.

|  | Propositions | NoC | Ground truth | Log$_{10}$*LR* HMC | EFM | STRmix™ |
|---|---|---|---|---|---|---|
| Case 1 | V+S01A | 2 | True donor | 20.15 | 20.18 | 20.15 |
| Case 2 | S02A+U+U | 3 | True donor | 17.03 | 17.28 | 16.98 |
| Case 2 | S02B+U+U | 3 | True donor | 7.50 | 7.88 | 7.26 |
| Case 2 | S02C+U+U | 3 | True donor | 5.41 | 6.11 | 5.83 |
| Case 2 | S02D+U+U | 3 | False donor | -16.18 | -2.36 | -14.03 |
| Case 3 | V+W+S03A | 3 | True donor | 7.87 | 6.82 | 7.69 |
| Case 3 | V+W+S03B | 3 | False donor | −∞ | −∞ | −∞ |
| Case 4 | V+S | 2 | True donor | 20.23 | 19.91 | 20.15 |
| Case 5* | S05A+U+U | 3 | True donor | 3.38 | 9.26 | 3.45 |
| Case 5 | S05B+U+U | 3 | True donor | 1.61 | 9.38 | 3.32 |
| Case 5 | S05C+U+U | 3 | False donor | -8.66 | **6.17** | -9.22 |

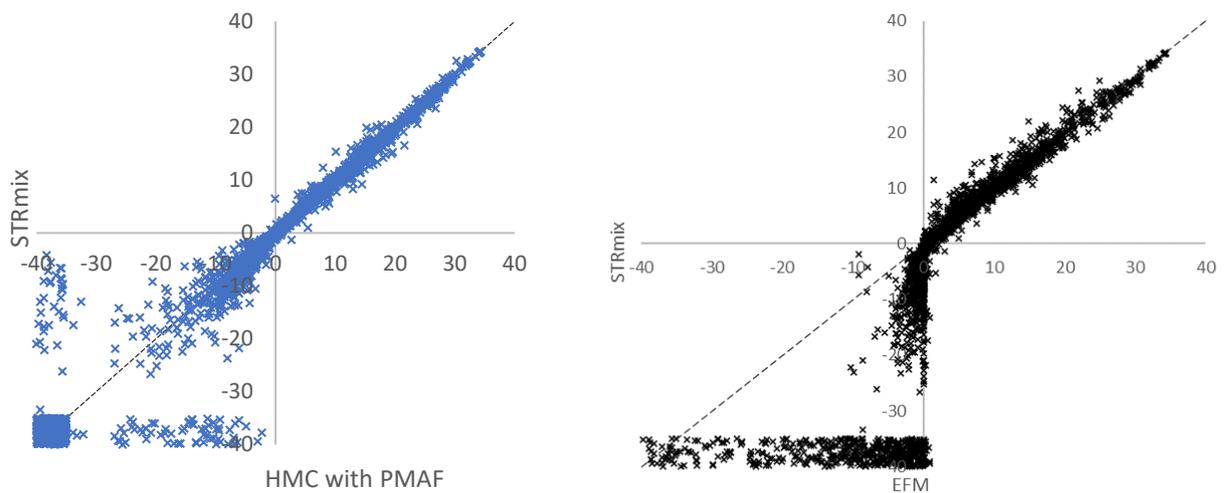

Figure 2: Log$_{10}$*LR*s assigned by HMC (left) with posterior mean allele frequency (PMAF) and EFM (right) vs log$_{10}$*LR*s assigned by STRmix™. Data for HMC with PMAF are taken from Susik and Sbalzarini [10]. EFM data are from are taken from Susik and Sbalzarini [10] for two- and three-contributor mixtures and Riman et al. [7] for four-contributor mixtures and hence the four-contributor mixtures will appear here and also in Figure 1**Error! Reference source not found.**. STRmix™ has its lowest precision for log$_{10}$*LR*s less than 0 and the scatter seen in the lower left quadrant of the left-hand plot is partly due to this. The much smaller scatter in the upper right-hand quadrant of the left-hand plot is affected in part by run to run variation inherent in the MCMC. This spread may be a reasonable indicator of our current precision for non-contributors.

Susik and Sbalzarini [10] provide data for HMC, EFM, and STRmix™ for a rework of the Riman et al. [7] data. They comment, correctly, that the limitation to two- and three-contributor mixtures for EFM is for reasons of analysis runtime, which were too large (regularly exceeding an hour per scenario) to be practical for the required numbers of repetitions on four-contributor mixtures. EFM v3.4.0 enables modelling of forward stutters. No model for double-backward stutters is available yet. The STRmix™ data are from Susik and Sbalzarini [10] but originate from Riman et al. [7]. We amend the result for two profiles[1] to the Riman et al. [7] extended analyses since we believe that these are closer to the correct analyses[2]. Susik et al. [10, 11] used the same input data used by Riman et al. [7]. They have independently verified that CleanIt has indeed filtered out some of the stutter peaks that are important for the analysis, confirming that this is an issue. However, they state that the Riman et al. [7] study was the only paper available at that time that makes it possible to compare with STRmix™ results with HMC. Therefore, all the three mentioned comparison papers (works of Riman et al., Costa et al. and Susik et al.) exhibit the same issue of missing stutter peaks. The comparison is given in Figure 2.

The Susik and Sbalzarini [10] statement about a lack of comparison data was correct at the time of their publication, but we have subsequently greatly enhanced the data available in the public domain (available at https://figshare.com/articles/dataset/ESR_response_to_NISTIR_8351_-_DRAFT_DNA_Mixture_Interpretation_A_NIST_Scientific_Foundation_Review/15062907?file=28980966) and continue that process in the supplementary material to this paper. This, we hope, meets the complaint made in the NIST Foundation review that there was insufficient data available to them by internet search [2]: KEY TAKEAWAY #4.3: Currently, there is not enough publicly available data to enable an external and independent assessment of the degree of reliability of DNA mixture interpretation practices, including the use of probabilistic genotyping software (PGS) systems.

Cheng et al. [16] also compare STRmix™ (V2.7.0) and EFM (V3.0.3). We reproduce some of their results in Figures 3 and 4.

---

[1] C02_RD14-0003-40_41-1;4-M2U15-0.315GF-Q1.1_03.15sec and H06_RD14-0003-48_49-1;4-M2e-0.315GF-Q1.0_08.15sec

[2] One of these Riman et al amendments is still not our preferred approach. This profile is affected by a known primer binding site mutation. Extending the MCMC chain does not amend the inappropriate modelling for this situation (see 15. Buckleton, J., et al., *Re: Riman et al. Examining performance and likelihood ratios for two likelihood ratio systems using the PROVEDIt dataset.* Forensic Science International: Genetics, 2022. **59**.)

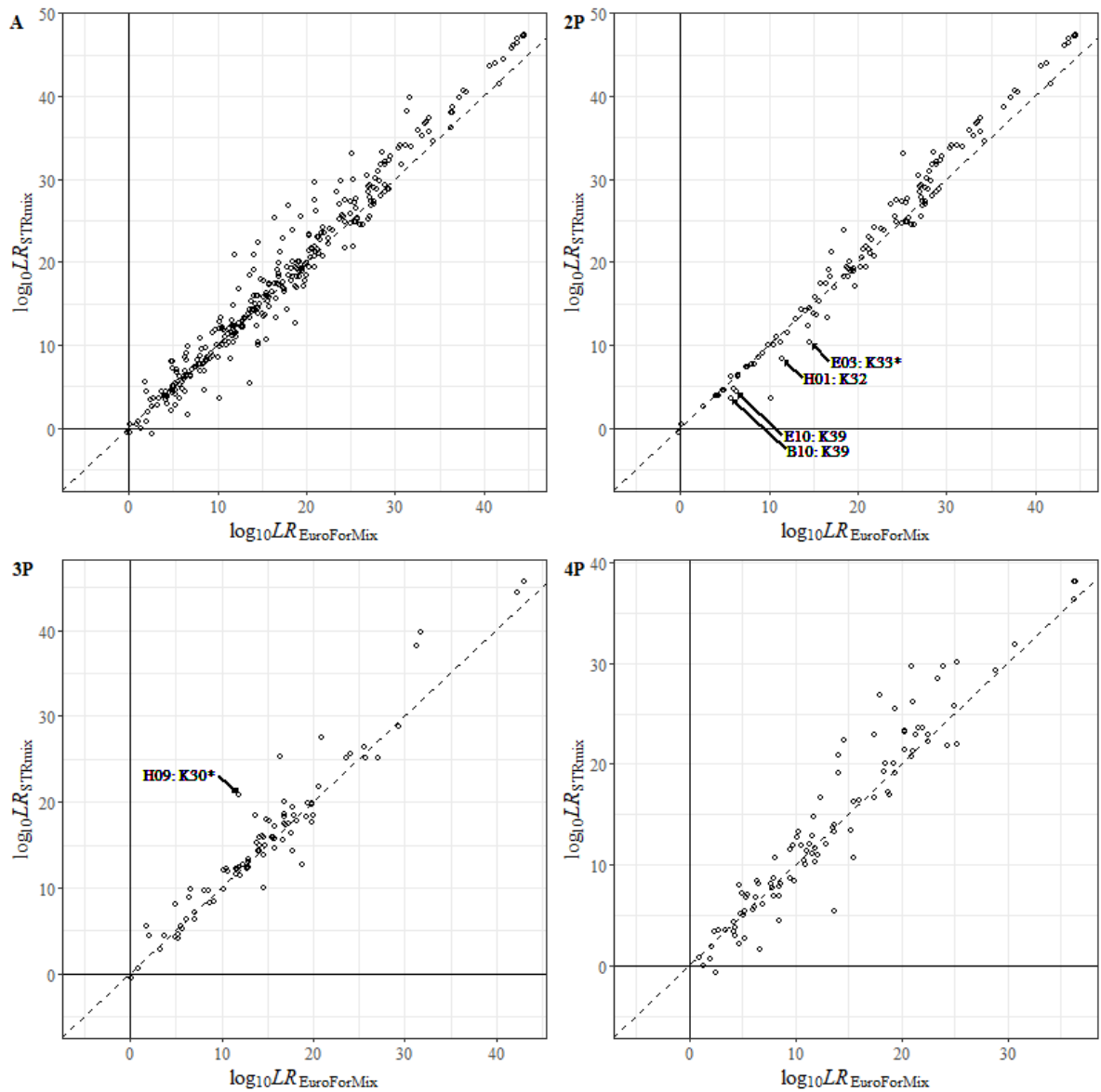

Figure 3: Scatter plot of the STRmix™ V2.7.0 $\log_{10}LR$ and EuroForMix V3.0.3 $\log_{10}LR$ for known contributors (circles) from the interpretation of all (panel A), two- (panel 2P), three- (panel 3P), and four-person (panel 4P) mixtures. The black arrows mark the five divergent results that were further investigated by Cheng et al. The label shows the sample identifier followed by the donor identifier. Reproduced from Cheng et al. [16] without any alteration for this publication.

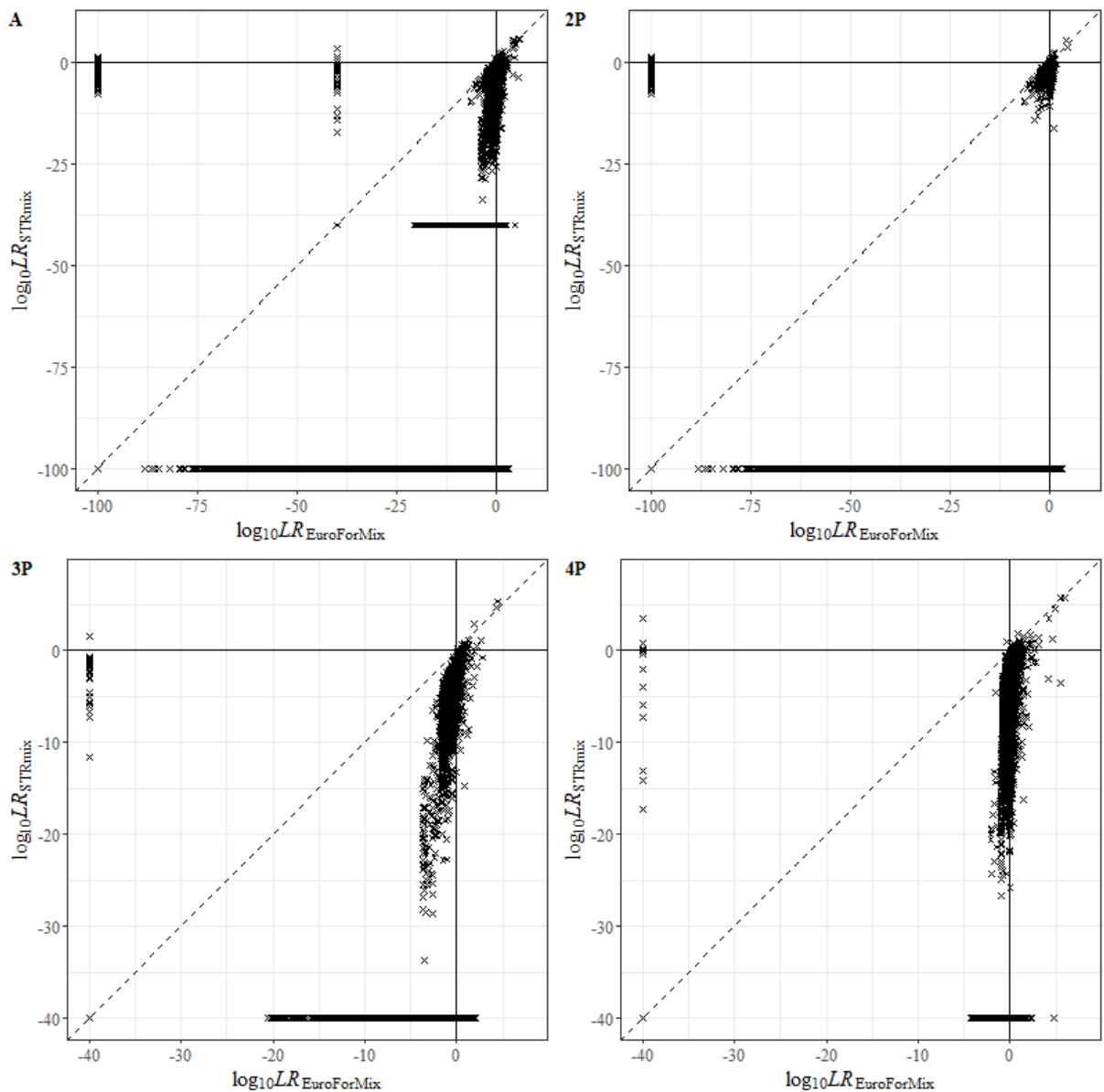

Figure 4: Scatter plot of the STRmix™ V2.7.0 $\log_{10}LR$ and EuroForMix V3.0.3 $\log_{10}LR$ for non-contributors (crosses) from the interpretation of all (panel A), two- (panel 2P), three- (panel 3P), and four-person (panel 4P) mixtures. *LR*s of 0 are presented as -100 for two-person mixtures, and -40 for three- and four-person mixtures. Reproduced from Cheng et al. [16] without any alteration for this publication.

EFM developers' advice is to always check the model validation results and not report the results if model validation failed and cannot be explained or corrected. We can confirm that this was done in Cheng et al. [16] but can neither confirm nor refute this for Costa et al. [17] or Riman et al. [7].

Riman et al. [7] do mention different models as a cause of difference in *LR*s between the PG software. However, they do not go further in determining in which way the models differ and how they cause differences. Cheng et al. [16] (see Figure 3), do investigate some of the large discrepancies in *LR* for known contributors between STRmix™ and EFM. Through their investigation they identify some key differences such as:

- minimum allele probabilities,

- stutter modelling, and,
- modelling separate locus specific amplification efficiencies.

The Cheng et al. [16] data are given in the supplementary material to this paper.

We briefly list below some of the known differences between STRmix™ and EFM.

### 1.2 *Peak height variance modelling*

STRmix™ introduces parameters for peak height variances. One of them expresses how much variance is allowed for allelic peak heights. Separate variance parameters are used for each stutter type modelled within the laboratory (for many laboratories this is at least back and forward stutter, and often double back, and half back at selected loci). The estimated distributions of these parameters are influenced by priors that are estimated based on empirical data (see [18] for an explanation of the method used to set these priors in STRmix™). This way the distributions should not diverge substantially from what has been observed in the past by the laboratory. EFM takes a different approach. One parameter is used to represent the peak height variances. This parameter is prior-less and set by the MLE. The value of the parameter can therefore move to larger values, under $H_p$, if the POI needs a larger variance for their genotype to better explain the observed profile.

Indeed, this can be observed in the results from Susik and Sbalzarini [10]. We present the variance parameters estimated by MLE within EFM for two- and three-contributor mixtures in Figure 5. Clearly, in many cases in which a non-contributor was the POI, the peak height variance is substantially larger (up to 3.34 times larger) in $H_p$ than in $H_d$ posterior estimation. This behaviour, we believe, leads to substantially higher *LR*s and higher frequency of *LR*>1 for non-contributors (see for example **Error! Reference source not found.**Figures 2 and 4).

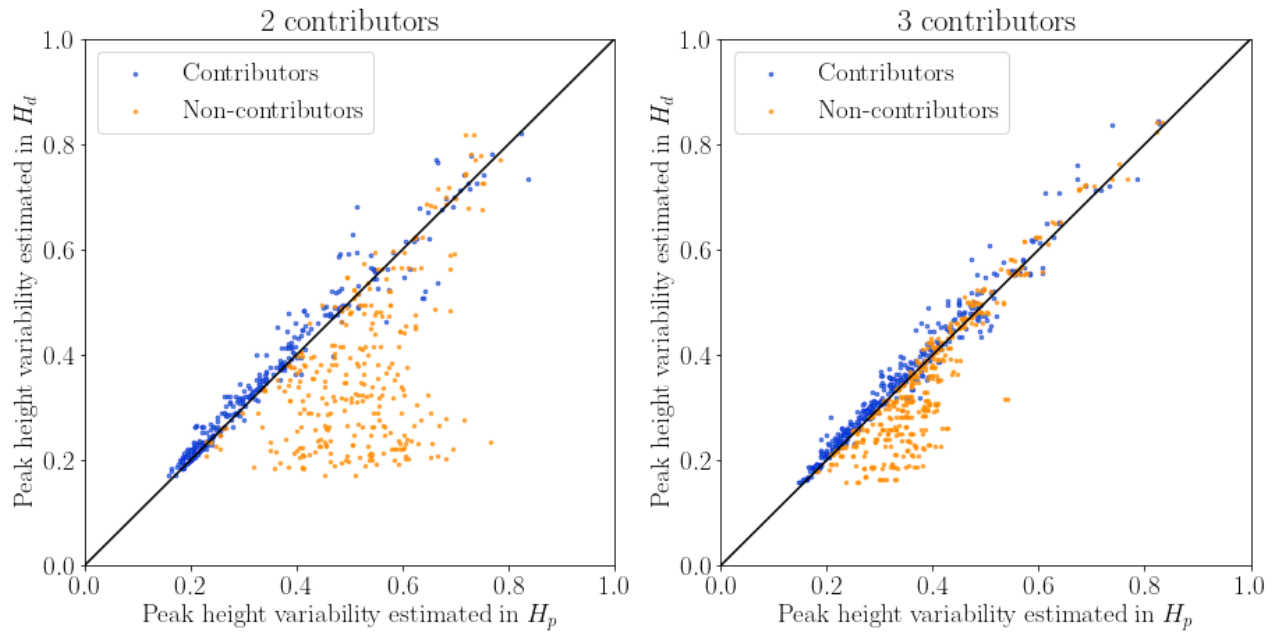

Figure 5: Comparison of values of parameters for the peak height variability in $H_p$ (*x*-axis) and $H_d$ (*y*-axis) estimated within EFM. The comparison is performed for two-contributor mixtures (left) and three-contributor mixtures (right). This is novel analysis by Susik of the run from Susik and Sbalzarini [10].

### 1.3 *Mixture proportions*

EFM estimates the mixture proportions separately under $H_p$ and the alternate proposition, $H_a$. In rare cases, this can result in significant differences in mixture proportion between the two propositions. For example, the datum bolded in **Error! Reference source not found.** gives the mixture proportions shown in **Error! Reference source not found.**.

Table 2: The mixture proportions from EFM (v1.11.4) for the NIST MIX13 Case 5 false donor 5C ($LR=4.95 \times 10^6$). This is interpreted as a three-person mixture although the ground truth is a four-person mixture because it is a perfect fit to three donors and cannot be blindly assigned as a four-person mixture.

| EFM mixture proportions | $H_p$ | $H_d$ | STRmix™ mixture proportions |
|---|---|---|---|
| Mix-prop. C1 | 0.3061 | 9.61E-11 | 0.51 |
| Mix-prop. C2 | 0.3469 | 0.3164 | 0.26 |
| Mix-prop. C3 | 0.3470 | 0.6836 | 0.23 |

It can be seen in **Error! Reference source not found.** that the solution forced in EFM under $H_p$ by the inclusion of non-donor 5C is close to 1:1:1 whereas the solution determined by MLE under $H_a$ is close to 2:1:0. Case 5 is a sample deliberately constructed from four donors chosen to have at most four alleles between them in the approximate ratio 1:1:1:1. The non-donor 5C is a hypothetical false donor constructed by taking alleles from different true donors.

### 1.4 *Stutter ratios*

STRmix™ assumes that every allele stutters differently and that the stutter ratio may be estimated empirically. EFM uses a single stutter proportion for all peaks. This proportion is estimated by the MLE and can be constrained with user-defined priors.

### 1.5 *Locus specific effects*

Apart from considering degradation, STRmix™ assumes that all loci can amplify slightly differently. Again, this difference is constrained and the prior on the extent of difference is set empirically. EFM assumes all loci amplify the same after considering degradation. This is demonstrably untrue and was shown to cause a problem by Cheng et al. [16] (see for example Figure 11A and Table 7 of Cheng et al. [16]).

### 1.6 *Minimum allele probabilities*

STRmix™ (V2.8 and above) assumes a previously unseen allele (unseen in the allele frequency database) has an unknown frequency modelled by $\beta(\frac{1}{k+1}, 2N)$ where $N$ is the number of individuals in the database and $k$ is the number of allelic classes. This method is based on standard Bayesian theory for updating the belief about a Binomial probability based on observed counts and the assumption of a Beta prior distribution however the mean of the resulting distribution is very small, $\frac{1}{k(2N+1)}$.

EFM assumes a previously unseen allele (unseen in the database) has exactly the frequency $\frac{5}{2N}$ where $N$ is the number of individuals in the database or alternatively may be set by the user. This is an empirical rule, rather well-grounded in theory – and is similar to the practice of `adding 5' to a table of counts so as to avoid issues around zero observations. As such, it is hard to justify on any other basis than it appears to be reasonably robust. Larger values are usually, but not always, conservative in mixture interpretation. This is a case where both models are correct in that they are both larger than the sample value which is clearly too low.

Susik and Sbalzarini [10] provide a comparison of the posterior allele probability model (STRmix™) vs the 5/2N model (HMC and EFM) by running their HMC software reusing the deconvolution provided by the inference model with both models. We reproduce their Figure 10 in Figure 6. Inspection of Figure 6 suggests that the rare allele probability model can account for some, but not all, of the differences noted between EFM and STRmix™.

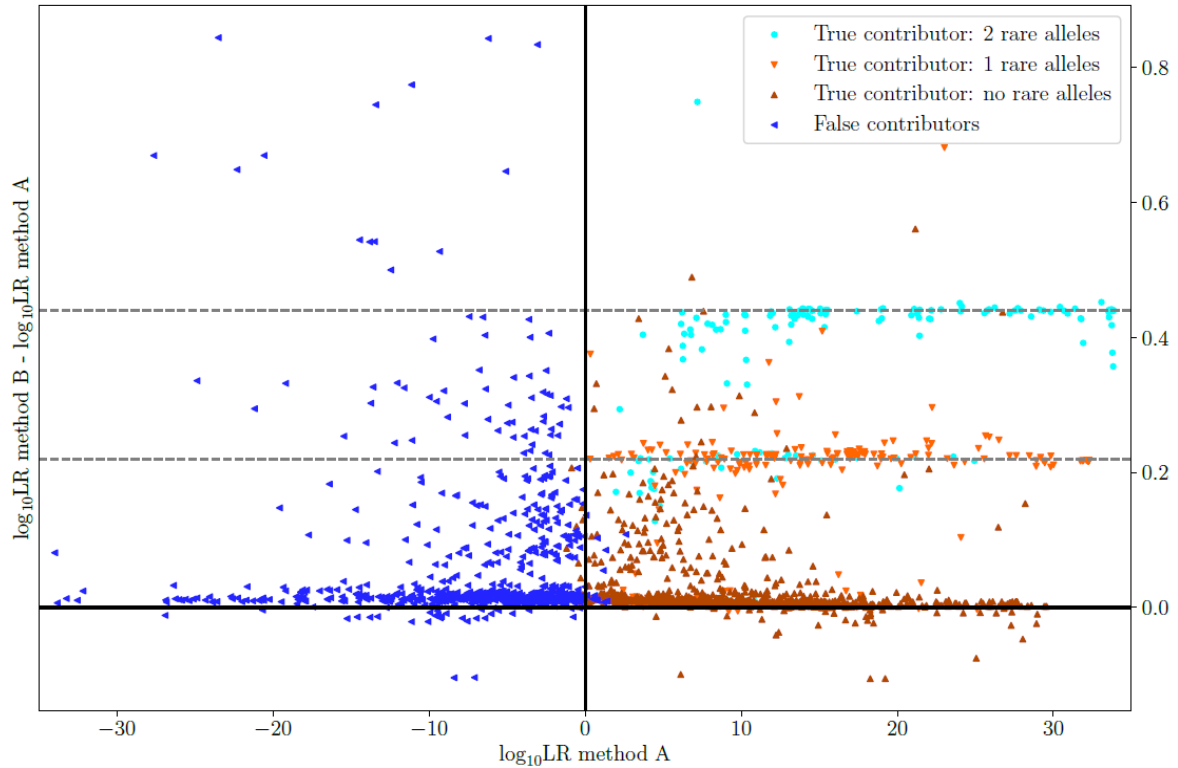

Figure 6: Difference between the $\log_{10}LR$ obtained with two different rare-allele models: the 5/2N model (Method A) and the posterior mean allele frequency model (Method B). A coancestry coefficient of 0.01 was used. Colours denote whether the POI is a true contributor and, if so, how many rare alleles are present in his/her genotype. The dashed horizontal lines are at differences of 0.22 and 0.44. The true POIs with rare alleles cluster around those lines. Reproduced from Susik and Sbalzarini [10] without any alteration for this publication.

At the risk of starting a deeply philosophical debate, all the comparison studies between EFM and STRmix™ have demonstrated that the two different models can both be useful. The most obvious practical example for this for EFM and STRmix™ is the minimum allele probability. This is discussed in Cheng et al. [16], but briefly both substitute a value conservative relative to the sample estimate for the allele probability when an allele is previously unobserved.

## 2.0 RESULTS AND DISCUSSION

STRmix™ determines a number proportional to the probability of the profile ($O$) for a genotype combination, $S_j$, that could explain the profile, $\Pr(O|S_j)$. It does this for all possible genotype combinations within an MCMC analysis, drawing values for model parameters from their posterior distribution. Post-MCMC, during a *LR* calculation, the nuisance parameters $S_j$ are summed across their prior and removed $\Pr(O) = \sum_j \Pr(O|S_j)\Pr(S_j)$. This is carried out conditioning on a proposition, which assigns some of the prior genotype set probabilities as 0.

In contrast, EFM uses maximum likelihood estimation (MLE) to separately optimise some of the model parameters (such as mixture proportion, stutter proportion, and importantly allele variance) under $H_p$ and $H_a$. By optimising the parameters separately under $H_p$ EFM allows them to move to "accommodate a POI" whether that POI is a true donor or not. This is most evident in **Error! Reference source not found.** for Case 5 propositions S05C+U+U where the log of the average *LR* for EFM was 6.17 for the non-donor S05C whereas HMC and STRmix™ both gave exclusionary *LR*s.

We note that MLE does not provide the required probability Pr(*O*) but rather a probability density $p(O|\hat{M})$ where $\hat{M}$ represents the modelling parameters that maximise the probability density of *O*. In the likely event that the modelling parameters do not have flat priors then this is not proportional to Pr(*O*). Even if the priors are flat the provided probability density is not proportional to Pr(*O*). This is further complicated if the number of dimensions of the sample spaces differ between $H_p$ and $H_a$, such as when the assigned number of contributors differs between the two propositions.

We provide a toy example in supplementary material 2 that demonstrates the difference between calculating a Bayes Factor through integration across the sample space and MLE. The comparison is made by considering an imbalanced DNA profile as originating from either two contributors (under $H_p$, as is required for a POI to be a contributor) or one contributor (under $H_a$).

The allele and stutter variances, and the stutter proportions cannot possibly have flat priors. For example, the values 0 and infinity cannot be possible for these variances and there must be a region of higher density somewhere around the expected values.

In EFM the parameters are the mixture proportions, peak height, degradation, stutter proportion, and the peak (used for both stutter and allele) variances. The parameters are estimated based on maximising the likelihood function (MLE). The event space of the desired probability is a volume in high dimensional space, where the number of dimensions is defined by the number of parameters and is not the same for different NoC. MLE (should) locate the highest density in this multidimensional space. However, it is the volume of this shape that is the probability needed. The height of the highest point (the MLE) is not equal to the volume desired and may not be proportional to it.

In order to have the same number of dimensions, one needs the same NoC in the numerator and denominator. Even with the same dimensionality there will be a difference in shape depending on the genotypes hypothesised to be present if one conditions on the POI in the separate estimation for $H_p$. We have not done a comprehensive survey, but we can find instances of both small and large differences between parameters' estimates. In **Error! Reference source not found.** we give some material from a typical EFM output.

Table 3. An excerpt of some of the outputs from EFM data from analyses in Cheng et al. [16].

| Sample ID. Both samples can be seen in **Error! Reference source not found.** | Example 1[3] | | Example 2[4] | |
|---|---|---|---|---|
| Estimates under | $H_p$ | $H_d$ | $H_p$ | $H_d$ |
| Param. | MLE | MLE | MLE | MLE |
| Mix-prop. C1 | 0.4812 | 0.68376 | 0.19376 | 0.3333 |
| Mix-prop. C2 | 0.5188 | 0.31624 | 0.40312 | 0.3333 |
| Mix-prop. C3 | | | 0.40312 | 0.3333 |
| P.H.expectation | 917.8 | 904.18 | 4068 | 4042.3 |
| P.H.variability[5] | 0.4992 | 0.44863 | 0.39506 | 0.46774 |
| Degrad. slope | 0.68826 | 0.69453 | 0.61877 | 0.62851 |
| BWstutt-prop. | 0.04488 | 0.029 | 0.10628 | 0.12246 |
| FWstutt-prop. | 2.79E-07 | 9.73E-09 | 1.01E-09 | 0.0183 |
| | | | | |
| logLik= | -476.8 | -503.3 | -885.1 | -912.4 |
| Lik= | 8.39E-208 | 2.59E-219 | 0 | 0 |

It is important to bear in mind that the outputs of STRmix™ and EFM are not probabilities. This affects any attempt to combine the outputs for different NoC. If they were probabilities this would be straightforward.

In STRmix™ the outputs are labelled $\Pr(E|H_p)$ and $\Pr(E|H_a)$. This is unfortunate. These are actually numbers proportional to the probability $\Pr(O|S_j)$. The constant of proportionality is the same for all genotype sets within a same NoC and hence will cancel within a *LR* calculation. This constant of proportionality is different for different NoC and hence one cannot simply combine $\Pr(E|H_p)$ for one NoC with $\Pr(E|H_a)$ for another and obtain an *LR*. This, and the solution, is discussed in Taylor et al. [20].

In EFM (when using MLE) the outputs are labelled under "estimates under Hp" and separately "estimates under Hd." The output is labelled "lik" for likelihood. These are neither probabilities, nor numbers directly proportional to them. The constant of proportionality is probably slightly different under $H_p$ and $H_a$ depending on how different the shape of the probability volume is. As stated, this difference appears to be variable.

These observations have important consequences when attempting to combine values across NoC. Slooten and Caliebe [21] made the insightful observation that the overall *LR* is the weighted average of *LR*s with the same number of contributors (NoC) under both propositions. The weights for this averaging involve both an assessment of the probability of the crime scene DNA profile and the probability of this NoC given the background information. Slooten and Caliebe rely on the assumption that the profile must have the same NoC under $H_p$ and $H_a$, but this number can range.

---

[3] H01_RD14-0003-31_32-1;1-M2c-0.062GF-Q2.0_08.25sec.hid for known contributor K32
[4] H09_RD14-0003-30_31_32-1;4;4-M2d-0.75GF-Q0.6_08.25sec.hid for known contributor K30
[5] This is one of two parameters that model the shape of the gamma distribution used for the peak height density curve. It is $\sigma$ in 19.    Bleka, Ø., G. Storvik, and P. Gill, *EuroForMix: An open source software based on a continuous model to evaluate STR DNA profiles from a mixture of contributors with artefacts.* Forensic Science International: Genetics, 2016. **21**: p. 35-44.

This is different from considering that the $H_p$ and $H_a$ can independently assign a NoC. Neither STRmix™ nor EFM give, directly, the probability of the crime scene DNA profile. The probability of this NoC given the background information is separate from an analysis of the profile.

2.1 *Meester and Slooten*

In this section we largely follow Meester and Slooten [8]. The following is the mathematical development of their argument for the EFM process. It can be omitted by the reader and the meaning will be retained. Consider:

$$LR_{ML} = \frac{p(O|H_1, G_P, \hat{M}_1, I)}{p(O|H_2, G_P, \hat{M}_2, I)} \quad \text{(equation 1.)}$$

Where:

- $LR_{ML}$ is the maximum likelihood $LR$
- $\hat{M}_1$ is the value of the mass parameters that maximises $p(O|H_1, G_P, \hat{M}_1, I)$
- $\hat{M}_2$ is the value of the mass parameters that maximises $p(O|H_2, G_P, \hat{M}_2, I)$
- $G_P$ is the genotype of the POI

Meester and Slooten [8] note that *"Despite our notation, the quotient in (equation 1.) is not a likelihood ratio, but instead it represents how much better the best explanation for the data under $H_1$ is than the best explanation under $H_2$."* This can be seen by noting that the numerator and denominator differ in two parameters (the $H$ terms and the $M$ terms) rather than one. In order for $LR_{ML}$ to be a likelihood ratio (for $H_1$ versus $H_2$), $\hat{M}_1 = \hat{M}_2 = \hat{M}$. However, it can be shown that $LR_{ML}$ is bounded by the likelihood ratios that assume the use of either $\hat{M}_2$ in both numerator and denominator, or $\hat{M}_1$. That is,

$$LR_{\hat{M}_2} = \frac{p(O|H_1, G_P, \hat{M}_2, I)}{p(O|H_2, G_P, \hat{M}_2, I)} \leq \frac{p(O|H_1, G_P, \hat{M}_1, I)}{p(O|H_2, G_P, \hat{M}_2, I)}$$
$$\leq \frac{p(O|H_1, G_P, \hat{M}_1, I)}{p(O|H_2, G_P, \hat{M}_1, I)} = LR_{\hat{M}_1} \quad \text{(equation 2.)}$$

This says that the quantity $LR_{ML}$ will be bounded somewhere between the $LR$s produced by optimising the parameter sets for one of the hypotheses. The ML estimates can be expected to give better estimates of the true $M$ (termed $M_0$ in our paper $M_0$) the more correct information that is considered. Hence $\hat{M}_1$ is better if $H_1$ is true and worse if it is false.

Meester and Slooten state that *"we can expect that $LR_{ML}(E)$ does not overstate the evidence for actual contributors and will be closer to the likelihood ratio at the true parameters $\theta_0$ (our $M_0$) for those, it will be anticonservative to do so for non-contributors."*

The Meester and Slooten prediction has been observed empirically. For example Cheng et al. [16], reproduced in **Error! Reference source not found.**, show many higher $LR$s for false donors in EFM than STRmix™. There is additional information in Cheng et al. [16] for example their Figure 10.

A comparison of STRmix™ V2.5.11 and EFM V1.10.0 on a database deliberately selected to increase the fraction of matching alleles appears in [14] and is reproduced in Figure 7.

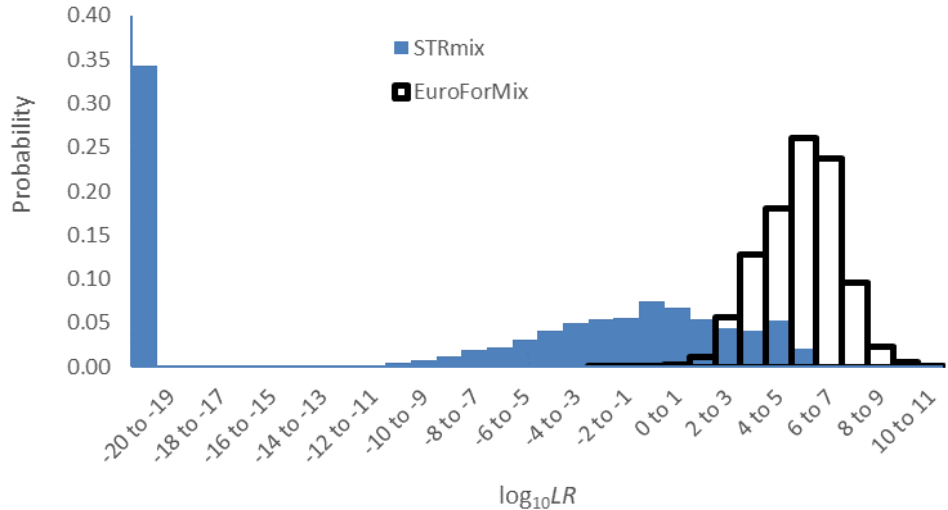

Figure 7: The results of 10,000 false donors tested against the NIST MIX13 interlaboratory study Case 5 profile using STRmix™ V2.5.11 and EuroForMix v1.10.0.

The non-contributors have been created by sampling, with replacement, from the alleles of the true donors. 41% of *LR*s are greater than 1 ($\log_{10}LR>0$) and 59% lower than 1 for STRmix™ 99.98% of *LR*s are greater than 1 for EuroForMix.

## 2.2 *Calibration*

The preferred way to examine the validity of a group of *LR*s is calibration [1, 22]. This is an empirical test to determine whether, on average, a group of *LR*s are approximately correct. The process is to take a set of $H_p$ and $H_d$ true tests. Since the number of $H_p$ and $H_a$ true tests are known then the prior odds are known. Using these prior odds, the posterior odds and hence the posterior probability of each analysis can be assigned. These are assembled into groups based on appropriately selected *LR* ranges. In each *LR* range there will be some analyses from $H_p$ true and some from $H_a$ true comparisons. The empirical frequency that an *LR* in this range is from a true donor can be calculated. This analysis is shown in **Error! Reference source not found.** and Figure 8 for EFM and STRmix™ using the data published in [16].

Table 4: The data for the calibration analysis from STRmix™ V2.7.0 and EuroForMix V3.0.3

| Log$_{10}$LR range | $H_p$ true counts | | $H_a$ true counts | | Expected Posterior probability | Observed Posterior frequency | |
|---|---|---|---|---|---|---|---|
| | STRmix | EFM | STRmix | EFM | | STRmix | EFM |
| 8 -9 | 15 | 12 | 0 | 0 | 0.999999-1 | 1 | 1 |
| 7-8 | 9 | 7 | 0 | 0 | 0.999991-0.999999 | 1 | 1 |
| 6-7 | 10 | 14 | 0 | 0 | 0.99991-0.999991 | 1 | 1 |
| 5-6 | 9 | 12 | 4 | 3 | 0.9991-0.99991 | 0.69 | 0.8 |
| 4-5 | 11 | 16 | 2 | 9 | 0.991-0.9991 | 0.85 | 0.64 |
| 3-4 | 12 | 3 | 3 | 3 | 0.914-0.991 | 0.8 | 0.5 |
| 2-3 | 4 | 5 | 2 | 27 | 0.514-0.914 | 0.67 | 0.16 |
| 1-2 | 2 | 3 | 18 | 151 | 0.096-0.514 | 0.1000 | 0.0195 |
| 0-1 | 4 | 3 | 82 | 2073 | 0.01-0.096 | 0.0465 | 0.0014 |
| -1-0 | 2 | 1 | 324 | 5714 | 0.0011-0.01 | 0.0061 | 0.0002 |
| -2 to -1 | 0 | 0 | 531 | 2488 | 0.0001-0.0011 | 0 | 0 |
| -3 to -2 | 0 | 0 | 578 | 1296 | 0.00001-0.0001 | 0 | 0 |
| -4 to -3 | 0 | 0 | 562 | 1431 | 0-0.00001 | 0 | 0 |
| Total across all *LR*s | 338 | 338 | 31912 | 31912 | | | |

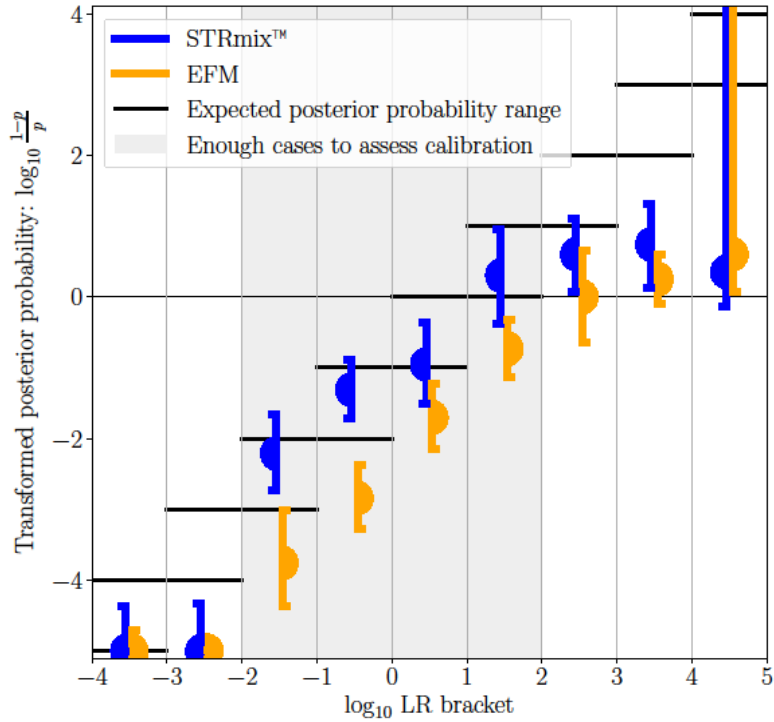

Figure 8: A plot of the binned calibration data given in **Error! Reference source not found.**. The error bars are assigned as 0.95 probability intervals. The horizontal lines represent the bounds on the expected posterior probability (logit transformed). We indicate with grey colour the area in which subjectively there is enough both of the $\log_{10}LR>0$ and $\log_{10}LR<0$ cases to assess calibration (see **Error! Reference source not found.** for the actual numbers).

An examination of Figure 8 shows a similar pattern for STRmix™ and EFM but with EFM tracking below STRmix™ and also below the expected bounds in the range $\log_{10}LR$ = -2 to +4. We cannot expect accuracy at high or low *LR*s in this plot because the counts of $H_a$ true at high and $H_p$ true at low *LR* are small. We believe that it is the separate optimisation of the parameters under $H_p$ that drives the low track of EFM between $\log_{10}LR$ = -2 to +4. The effect of this separate optimisation is increased by the unconstrained parameters in EFM.

### 3.0 CONCLUSIONS

The best path to finding reliable systems is calibration.

The separate optimisation of the parameters assuming the presence of the POI combined with the use of MLE was hypothesised by Meester and Slooten [8] to cause higher *LR*s for non-donors. This is observed in multiple empirical tests reprised here and is the probable cause of EFM's assignment of higher *LR*s for non-donors. In the main, the difference between EFM and STRmix™'s *LR*s are between degrees of support for exclusion or low inclusionary support. On the positive side this behaviour does buffer EFM against false exclusions caused by input file errors, genetic anomalies, and extreme PCR events. However, it is also likely that this assignment of high *LR*s for non-donors is driving behavior such as non-reporting of *LR*s below $10^6$ or $10^4$ to the detriment of falsely implicated non-donors.

**Supplementary Material 2: Example of variable NoC in $H_1$ and $H_2$ using integration vs MLE**

Consider a simplified example where there are two peaks at one locus. These are at allele A of height 1000 rfu and at allele B at height 1100 rfu. This is a hypothetical perfect multiplex which has no stutter or degradation, simply for simplicity.

Let:

- $G_P = [B,B]$
- $H_1$: The DNA came from the POI and an unknown person unrelated to the POI
- $H_2$: The DNA came from an unknown person unrelated to the POI
- $c^2 = 12$
- $T_i$: the template of contributor $i$

We consider, as seems reasonable from these peak heights, that the unknown person must be genotype AB. For this toy example let the observed peak heights be $O_i$ and the expected peak heights $E_i$. Let the peak heights be independent given the template $T$, and the genotypes of the contributors, $S_j$.

We model $p(\log \frac{O_i}{E_i}) \sim N\left[0, \frac{c^2}{E_i}\right]$ and $E_i = \begin{cases} \sum_i t_i & i \in S_j \\ 0 & i \notin S_j \end{cases}$

The $LR$ using maximum likelihood is

$$LR_{ML} = \frac{p(O \mid H_1, S_j, \hat{T}_1, \hat{T}_2) \Pr(S_j \mid H_1)}{p(O \mid H_2, S_j, \hat{T}_1) \Pr(S_j \mid H_2)}$$

$$= \frac{p(O \mid H_1, S_j, \hat{T}_1, \hat{T}_2)}{p(O \mid H_2, S_j, \hat{T}_1)}$$

where $\hat{T}_1, \hat{T}_2$ are the MLE estimate for the $T_i$

The $LR$ using integrated probabilities is

$$LR_{int} = \frac{\Pr(S_j \mid H_1) \int_{T_1} \int_{T_2} p(O \mid H_1, S_j, T_1 = t, T_2 = t') p(T_1 = t) p(T_2 = t') dT_1 dT_2}{\Pr(S_j \mid H_2) \int_{T_1} p(O \mid H_2, S_j, T_1 = t) p(T_1 = t) dT_1}$$

$$= \frac{\int_{T_1} \int_{T_2} p(O \mid H_1, S_j, T_1 = t, T_2 = t') p(T_1 = t) p(T_2 = t') dT_1 dT_2}{\int_{T_1} p(O \mid H_2, S_j, T_1 = t) p(T_1 = t) dT_1}$$

In Figure 9 we give a section of a spreadsheet that implements the model described above. The values in this figure are $p(O \mid H, S_j, T_1 = t, T_2 = t')$. The MLE estimates are given in Table 3 and can be located in Figure 9 (they are bolded). The integral needs a prior on template and we model that here as U[0,30,000] for each contributor. These values appear in Table 3. The higher order proposition should always have a higher (or equal) MLE and does so in this example. Therefore, the $LR$ based on MLE should always favor the higher order proposition. The integral for the NoC = 2 solution ($H_1$) is much lower. This occurs because the peak is higher but vast parts of the full volume have low density.

Table 3. The MLE, integral, and LR estimates for the toy example described.

|          | $H_1$  | $H_2$  | LR     |
|----------|--------|--------|--------|
| MLE      | 13.59  | 14.12  | 1.04   |
| integral | 0.2051 | 0.0018 | 0.0088 |

|  | t2 | | | | | | |
| --- | --- | --- | --- | --- | --- | --- | --- |
| t1 | 0 | 50 | 100 | 150 | 200 | 250 | 300 |
| 275 | 0.00 | 0.00 | 0.01 | 0.02 | 0.04 | 0.08 | 0.12 |
| 325 | 0.00 | 0.01 | 0.02 | 0.05 | 0.09 | 0.16 | 0.23 |
| 375 | 0.01 | 0.02 | 0.05 | 0.11 | 0.20 | 0.31 | 0.41 |
| 425 | 0.02 | 0.06 | 0.12 | 0.23 | 0.39 | 0.57 | 0.72 |
| 475 | 0.06 | 0.13 | 0.27 | 0.48 | 0.75 | 1.01 | 1.18 |
| 525 | 0.13 | 0.28 | 0.55 | 0.92 | 1.33 | 1.68 | 1.82 |
| 575 | 0.28 | 0.58 | 1.05 | 1.64 | 2.22 | 2.60 | 2.62 |
| 625 | 0.58 | 1.12 | 1.88 | 2.74 | 3.44 | 3.74 | 3.50 |
| 675 | 1.11 | 2.00 | 3.13 | 4.24 | 4.95 | 5.00 | 4.36 |
| 725 | 1.98 | 3.33 | 4.84 | 6.09 | 6.62 | 6.20 | 5.03 |
| 775 | 3.28 | 5.13 | 6.95 | 8.12 | 8.19 | 7.14 | 5.38 |
| 825 | 5.03 | 7.33 | 9.22 | 10.01 | 9.39 | 7.61 | 5.34 |
| 875 | 7.17 | 9.70 | 11.34 | 11.44 | 9.97 | 7.52 | 4.92 |
| 925 | 9.44 | 11.88 | 12.90 | 12.09 | 9.80 | 6.88 | 4.20 |
| 975 | 11.52 | 13.46 | 13.58 | 11.84 | 8.93 | 5.84 | 3.32 |
| 1025 | 13.01 | **14.12** | 13.24 | 10.73 | 7.53 | 4.59 | 2.44 |
| 1075 | **13.59** | 13.71 | 11.95 | 9.01 | 5.89 | 3.35 | 1.66 |
| 1125 | 13.15 | 12.33 | 9.99 | 7.01 | 4.28 | 2.27 | 1.05 |
| 1175 | 11.78 | 10.27 | 7.74 | 5.06 | 2.88 | 1.43 | 0.62 |
| 1225 | 9.78 | 7.93 | 5.57 | 3.39 | 1.80 | 0.83 | 0.34 |
| 1275 | 7.52 | 5.68 | 3.71 | 2.11 | 1.05 | 0.45 | 0.17 |
| 1325 | 5.37 | 3.77 | 2.30 | 1.22 | 0.56 | 0.23 | 0.08 |
| 1375 | 3.56 | 2.33 | 1.32 | 0.66 | 0.28 | 0.11 | 0.04 |
| 1425 | 2.19 | 1.33 | 0.71 | 0.33 | 0.13 | 0.05 | 0.01 |
| 1475 | 1.25 | 0.71 | 0.35 | 0.15 | 0.06 | 0.02 | 0.01 |
| 1525 | 0.67 | 0.35 | 0.16 | 0.07 | 0.02 | 0.01 | 0.00 |
| 1575 | 0.33 | 0.16 | 0.07 | 0.03 | 0.01 | 0.00 | 0.00 |
| 1625 | 0.15 | 0.07 | 0.03 | 0.01 | 0.00 | 0.00 | 0.00 |
| 1675 | 0.07 | 0.03 | 0.01 | 0.00 | 0.00 | 0.00 | 0.00 |
| 1725 | 0.03 | 0.01 | 0.00 | 0.00 | 0.00 | 0.00 | 0.00 |

Figure 9: Values for $p(O|H,S_j,T_1=t,T_2=t')$ plotted against $T_1$ and $T_2$ for the toy example described